\documentstyle[preprint,aps,psfig]{revtex}
\tightenlines
\def\beq{\begin{equation}}
\def\eeq{\end{equation}}
\def\bea{\begin{eqnarray}}
\def\eea{\end{eqnarray}}
\def\nnu{\nonumber}
\def\tst{\textstyle}

\def\fno#1{Fig.~\ref{#1}}
\def\cno#1{\cite{#1}}
\def\eno#1{Eq.~(\ref{#1})}
\def\etwo#1#2{Eqs.~(\ref{#1}) and (\ref{#2})}
\def\rno#1{Ref.~\cite{#1}}

\def\gam{\gamma}
\def\dta{\delta}

\def\lam{\lambda}

\def\apx{\approx}

\def\hf{{1\over2}}
\def\tshf{\tst\hf}

\def\lp{\left(}
\def\rp{\right)}

\def\ham{{\cal H}}
\def\ket#1{|#1\rangle}

\def\tran#1#2{\langle#1|#2\rangle}

\def\mel#1#2#3{\langle#1|#2|#3\rangle}

\def\bH{{\bf H}}
\def\bJ{{\bf J}}

\def\bu{{\bf u}}

\def\xhat{\bf{\hat x}}
\def\yhat{\bf{\hat y}}
\def\zhat{\bf{\hat z}}

\def\Fe8{Fe$_8$}
\def\Jzx{\lp \begin{array}{c} J_z \\ J_x \end{array} \rp}
\def\bla{\bar\lam}
\def\sla{\sqrt{\lam}}
\def\sbl{\sqrt{\bla}}
\def\slbl{\sqrt{\lam\bla}}
\def\bham{\overline\ham}

\begin{document}
\draft

\title{Diabolical Points
in Magnetic Molecules: An Exactly Solvable Model}

\author{Ersin Ke\c{c}ecio\u{g}lu and Anupam Garg$^*$}
\address{Department of Physics and Astronomy, Northwestern University,
Evanston, Illinois 60208}

\date{\today}

\maketitle

\begin{abstract}
The magnetic molecule \Fe8 has been observed to have a rich pattern
of degeneracies in its magnetic spectrum as the static magnetic field
applied to the molecule is varied. The points of
degeneracy, or diabolical points in the magnetic field space, are
found exactly in the simplest model Hamiltonian for this molecule.
The points are shown to form a perfect centered rectangular lattice,
and are shown to be multiply diabolical in general. The multiplicity
is found. An earlier semiclassical solution to this problem is thereby
shown to be exact in leading order in $1/J$ where $J$ is the spin.
\end{abstract}

\widetext

A famous theorem of quantum mechanics states that the intersection
of two energy levels of a physical system is infinitely unlikely
as a single parameter is varied \cno{vnw}. Instead, level repulsion
is the rule, and level crossing happens only when there is some
symmetry. If one can vary more than one parameter, however
(at least two if the Hamiltonian is real, three if it is complex),
then degeneracies can be found at isolated points in the parameter
space \cite{ll,hlh,bw,va,fn1}. Such degeneracies are usually called
``accidental", for as a rule the points obey no discernible pattern.
Berry and Wilkinson's coinage for such degeneracies, ``diabolical,"
emphasizes the same point, although the term originates
less in any desire to attribute the phenomenon to a Mephistophelean
hand, than in the fact that the energy surface near the degeneracy
is a double cone, like an Italian toy called the {\it diavolo}.
When exceptions occur, and the pattern of diabolical points or
degeneracies is not random, as in the Kepler problem, or in the
isotropic harmonic oscillator, one seeks (and sometimes finds)
a higher dynamical symmetry of the Hamiltonian. 

It is therefore of some interest that a whole array of diabolical
points has been discovered by Wernsdorfer and Sessoli (WS) \cno{ws}
in the magnetic spectrum of the
molecular solid [(tacn)$_6$Fe$_8$O$_2$(OH)$_{12}$]$^{8+}$ (shortened
to \Fe8 from now on). Strong intramolecular exchange interactions
among the Fe$^{3+}$ ions lead to a total spin of $J=10$ in the ground
manifold for each molecule, which can therefore be conceived of
as a single large spin of this magnitude. The dipolar interactions
between molecules are weak, and may be ignored. Spin-orbit
interactions lead to an easy-axis anisotropy, with further
differentiation in the non-easy plane.  The simplest anisotropy
Hamiltonian that describes the magnetic properties of one molecule is
\beq
\ham  = -k_2 J_z^2 + (k_1 - k_2) J_x^2 - g\mu_B \bJ\cdot\bH, \label{ham}
\eeq
with $k_1 > k_2 > 0$ ( $k_1 \apx 0.33$~K, and $k_2 \apx 0.22$~K
\cno{fot}), $g = 2$, and $\bH$ is an external magnetic field. With
\eno{ham}, the easy, medium, and hard axes are $\zhat$, $\yhat$, and
$\xhat$, respectively.

To understand the evidence for diabolicity, let us suppose that
the field $\bH$ is along the hard axis, $\xhat$, and not too large.
The easy directions, which are along $\pm\zhat$ when $\bH = 0$, cant 
symmetrically toward $\xhat$. At first sight, as $H_x$ is increased, it
should be progressively
easier for the spin to tunnel from a state localized in the potential
well in the positive $J_z$ hemisphere to the symmetrically located
state in the negative hemisphere.  It was found some time ago,
however, that in the model (\ref{ham}), the tunnel
splitting between the ground
states in the positive and negative $J_z$ wells actually
oscillates as a function of $H_x$, going to zero
at periodically spaced $H_x$ values \cno{epl1}. Precisely these
oscillations have been seen by WS in \Fe8.
The oscillations were initially explained \cno{epl1}
in terms of an interference
between instanton trajectories for the spin, and the observations on
\Fe8 are very hard to explain by other mechanisms. The clincher is that
WS see additional oscillations that were not
predicted. To understand these, suppose $\bH$ also has a nonzero
$z$ component so as to bring the first or second excited state in the
positive $J_z$ well into degeneracy with the ground state in the other
well. One can then
conceive of tunneling between these states. WS observe that the amplitude
for this tunneling also oscillates with $H_x$, and that the oscillations
are shifted by half a period for each excited state in the deeper well. 

That the tunnel splitting between two states vanishes, is merely another
way of saying that the states are exactly degenerate. The degeneracies
that lie on the $H_x$ or $H_z$ axes in the magnetic field space can be
understood in terms of symmetry allowed level crossings as per the von
Neumann-Wigner theorem \cno{prb1}, but the newly discovered ones by WS
lie off the axes, and cannot be so understood. They are truly nontrivial
instances of diabolical points. Since the experiments were reported,
they have been successfully explained by one of us (AG) \cno{prletc},
and independently, Villain and Fort \cno{vf}. The approach is based
on a discrete version of the phase integral or
Wentzel-Kramers-Brillouin method, and is semiclassical in nature,
with $1/J$ playing the role of $\hbar$ in continuum problems with
massive particles. It is found that the 
$\ell'$th level in the negative $J_z$ well (with $\ell'=0$
being the lowest level) and the $\ell''$th level in the positive
one are degenerate when $H_y = 0$, and (see \fno{dpts})
\bea
{H_z(\ell',\ell'') \over H_c}
    &=& {\sla (\ell'' - \ell') \over 2 J}
                 \label{plhz} \\
{H_x(\ell',\ell'') \over H_c}
   &=& {\sqrt{1-\lam} \over J}
        \left[ J - n - \tshf (\ell' + \ell'' + 1) \right],
	     \label{plhx}
\eea
with $n = 0, 1, \ldots, 2J - (\ell' + \ell'' + 1)$. Here,
$\lam = k_2/k_1$, and $H_c = 2 k_1 J/g\mu_B$. Note that these
equations do give a half-period shift per excited state, as seen
by WS.

The surprise is that although \etwo{plhz}{plhx} are
dervied semiclassically, and should have higher order corrections
in $1/J$, they appear to be exact as written!
This has been noted by both
Villain and Fort, and AG. The evidence is from (a) analytic
diagonalization for $J \le 2$, (b) perturbation theory in
$\lam$ \cno{epl2}, and (c) numerics. Note that if exact,
\etwo{plhz}{plhx} would imply not only that the diabolical points lie
on a perfect centered rectangular lattice in the $H_x$-$H_z$
plane, but also that many of the points are {\it multiply}
diabolical, i.e., that more than one pair of levels is simultaneously
degenerate. It is easily shown that the multiplicity is as indicated
in \fno{dpts}: If we arrange the points into concentric rhombi,
those on the outermost rhombus are singly diabolical (i.e., there
is only one pair of degenerate states), those on the next rhombus
are doubly diabolical (two pairs of degenerate states), and so on.

In this paper we shall prove that this perfect lattice hypothesis
is in fact true. Not only is this an interesting problem in
mathematical physics in its own right, but we believe that it will
help understand real \Fe8 also. Exact solutions generally open the way
for perturbative treatment of small corrections, and in \Fe8, we beleive
our work will enable us to better treat such effects as the higher order
anisotropies and the dipolar interactions mentioned above, and allow
more detailed understanding of nonzero temperature effects \cno{ws2}.

Before giving our proof, we should note that we have not been able
to find if the Hamiltonian (\ref{ham}) has a higher symmetry at the
diabolical points. Such a suspicion is natural, given the experience
with the exceptional cases mentioned in our first paragraph.
Likewise, we have not succeeded in finding the wavefunctions. We
also note that the semiclassical approximation is demonstrably
inexact at non diabolical values of $\bH$, in order to dispel any
suspicion in the readers' minds that the classical and quantum
dynamics of the Hamiltonian (\ref{ham}) are identical as in the
case of the harmonic oscillator.

We now present our proof. It proceeds in three steps. In step
1 we perform a spin rotation about $\yhat$ so that $\ham$ no longer 
has any terms in $J_x^2$. The Hamiltonian is then tridiagonal in the
new $J_z$ basis. In step 2 we make use of a necessary
condition for a tridiagonal Hermitean matrix to have degenerate
eigenvalues, and thus determine the possible locations of any diabolical
points. In step 3, we use continuity and topological arguments to
find the multiplicity of each of these diabolical points. The results
are precisely those given above.

{\it Step 1:} Let us rotate $\bJ$ about the $\yhat$ axis so that
\beq
\Jzx \to \lp \begin{array}{cc}
              \sla    &   \sbl \\
             -\sbl    &   \sla
             \end{array} \rp \Jzx,  \label{rot}
\eeq
where $\bla = 1-\lam$. It is also convenient to define scaled fields
$u_x$, $u_y$, and $u_z$ via
\beq
\bH = {H_c \over 2J}(
       \begin{array}{lcr}
        \!\sbl u_x, & u_y, & \sla u_z\!
        \end{array}), \label{defu}
\eeq
to scale all energies by $k_1$, and to write $\bham = \ham/k_1$.
In the new axes, we have
\bea
\bham &=& -(\lam - \bla)J_z^2 - \slbl(J_z J_x + J_x J_z) \nnu \\
        &&\quad -\left[ (\lam u_z - \bla u_x)J_z
               +\slbl(u_x + u_z) J_x + u_y J_y\right]. \label{Htri}
\eea

{\it Step 2:} The Hamiltonian (\ref{Htri}) is tridiagonal in the
the new $J_z$ basis, $J_z \ket m = m\ket m$.
Its only nonzero matrix elements are
$\mel{m}{\bham}{m} \equiv w_m$, and $\mel{m}{\bham}{m'} \equiv t_{m,m'}$,
with $m' = m \pm 1$. And, it is obviously Hermitean:
$w_m^* = w_m$, $t_{m,m'} = t^*_{m',m}$. For such matrices, it is a
theorem that all the eigenvalues are simple (i.e., nondegenerate) if
none of the $t_{m,m'}$ vanish \cno{sb}. This result
is physically almost obvious if one thinks of $\bham$ as a tight-binding
model for an electron in one dimension with on-site energies $w_m$
and nearest neighbor hopping elements $t_{m,m\pm 1}$. Two states which
were degenerate could be spatially localized in different regions of the
lattice. But then, since $t_{m,m\pm 1} \ne 0$ for any $m$, it would be
possible for the electron to hop from one region to the other, which is
self-contradictory. The rigorous proof by Stoer and Bulirsch consists
of noting that the successive diagonal subdeterminants of $\bham - E I$
(where $I$ is the unit matrix), form a Sturmian sequence of polynomials
$p_j(E)$, $j = 1, 2, \ldots, 2J+1$, with the properties that $p_j$ is of
degree $j$, has $j$ real roots, each one of which is simple, and is
strictly bracketed by two roots of $p_{j+1}$.

It follows that the diabolical points of $\bham$, if any, must
lie on the loci in magnetic field space defined by $t_{m,m+1} = 0$. Using
the standard representation of the angular momentum matrices, we have
\beq
t_{m,m+1} = -\hf \left[
           \lp u_x + u_z + (2m+1)\rp\slbl + i u_y \right]
           [J(J+1) - m(m+1)]^{1/2}. \label{tmmp}
\eeq
The real and imaginary parts of this quantity must vanish separately
for some $m$. We thus conclude that any diabolical points must lie
in the $H_x$-$H_z$ plane:
\beq
u_y = 0, \label{uy0}
\eeq
and in this plane on the lines
\beq
u_x + u_z = -(2m_0+1),\quad m_0 = -J, -J+1, \ldots, J-1. 
    \label{uxz+}
\eeq
Further, when these conditions are obeyed, $\bham$ divides into
two blocks, of size $n_n$ and $n_p$, with $n_n = J + m_0 + 1$,
and $n_p = J - m_0$. In the first block, $m \le m_0$, and
in the second $m \ge m_0 + 1$.

The conditions (\ref{uy0}) and (\ref{uxz+}) are not enough to locate
the diabolical points. However, we can repeat our
argument, starting with a rotation (\ref{rot}) in which the signs
of the $\sbl$ entries are reversed. In this way we find that at a
diabolical point, we must also obey the condition
\beq
u_x - u_z = -(2n_0+1),\quad n_0 = -J, -J+1, \ldots, J-1,
    \label{uxz-}
\eeq
which along with \eno{uxz+} fixes the points completely.
Solving these equations, we get
\bea
\begin{array}{rcl}
u_x &=& -(m_0 + n_0 + 1),\\
u_z &=& n_0 - m_0,
\end{array}        \label{uxuz}
\eea
which with \eno{defu} are identical to \etwo{plhz}{plhx}.

{\it Step 3:} It remains to find just how many pairs of levels
are degenerate at each of the points (\ref{uxuz}). To do this, we
first show that under a continuous change in the parameter $\lam$,
the diabolical points of the Hamiltonian (\ref{ham}) must evolve
smoothly in the $u_x$-$u_z$ plane. In particular, the number of
diabolical points must stay fixed. We present two arguments for
this, one more physical, and the other more mathematical.

The physical argument is much like the standard one
\cno{ll} for the von Neumann-Wigner theorem. Suppose
that for some $\lam = \lam_0$, $\bham$ has two degenerate states
$\ket{\psi_a}$ and $\ket{\psi_b}$ at $u_x = u_{x0}$, $u_z = u_{z0}$.
(We set $u_y = 0$ throughout.) Now let $\lam$ be changed by a small
amount $\dta\lam$. We seek new eigenstates in the first approximation
as linear combinations of $\ket{\psi_a}$ and $\ket{\psi_b}$. The
secular matrix in this subspace is given by
\beq
\lp \begin{array}{cc}
\bham_{aa} & \bham_{ab} \\
\bham_{ba} & \bham_{bb} 
\end{array} \rp,  \label{secu}
\eeq
with $\bham_{ba} = \bham_{ab}$ since with $u_y = 0$,
the Hamiltonian is real.
Here, $\bham_{aa}$, $\bham_{ab}$, etc., are all smooth functions of
$\lam$, $u_x$, and $u_z$. The eigenvalues of this matrix will be equal
if and only if  $\bham_{aa} = \bham_{bb}$, and $\bham_{ab} = 0$.
Expanding these conditions in the deviations from $\lam_0$, $u_{x0}$,
and $u_{z0}$, we get
\beq
\lp \begin{array}{cc} a & b \\ c & d  \end{array} \rp
\lp \begin{array}{c} \dta u_x \\ \dta u_z \end{array}\rp
=
\dta\lam \lp \begin{array}{c}   e \\ f \end{array}\rp,
  \label{smalldev}
\eeq
where the quantities $a$ through $d$ are partial derivatives of
$\bham_{aa}$ etc. with respect to $u_x$ and $u_z$, and $e$ and $f$ are
partial derivatives with respect to $\lam$. Since the diabolical
point for $\lam = \lam_0$ is isolated, it follows that the only
solution to \eno{smalldev} when $\dta\lam = 0$ is
$\dta u_x = \dta u_z = 0$, i.e., that the matrix on the left hand side
is nonsingular, and a nonzero solution can be found when $\dta\lam \ne 0$.
It follows in turn that the diabolical point can only be moved, and
not eliminated or created by changing $\lam$.

The second argument is based on Berry's phase \cno{mvb}. Let us
consider a diabolical point as in the preceding paragraph, and let
$C$ be a small closed contour in the $u_x$-$u_z$ plane encircling
this point. Berry's phase is given by
\beq
\gam(C) = i \oint_C
   \tran{\psi_a(\lam,\bu)}{\nabla_{\bu}\psi_a(\lam,\bu)}\cdot d\bu,
         \label{berph}
\eeq
where $\bu = (u_x,u_z)$, and $\nabla_{\bu}$ is a gradient 
with respect to these fields. As shown by Berry, $\gam(C) = \pm\pi$
if $C$ encloses a true diabolical point, and $\gam(C) = 0$ if the
two states merely approach each other very closely without ever being
degenerate.
[Actually, since our Hamiltonian is real, and the parameter space
$(u_x,u_z)$ is two-dimensional, we really only need the weaker
result due to Herzberg and Longuet-Higgins \cno{hlh} for the
sign change of the wavefunction upon encircling the
degeneracy: $e^{i\gam(C)} = -1$.] Since the perturbation $\dta\bham$
engendered by changing $\lam$ or $\bu$ by a small amount is non-singular,
$\ket{\psi_a(\lam,\bu)}$ is a smooth function of $\lam$ and $\bu$. It
follows that if $\lam$ varies continuously, the integrand of
\eno{berph} can not change discontinuously. Hence, for small enough
$\dta\lam$, the phase $\gam(C)$
must continue to be what it was for $\lam = \lam_0$, $+\pi$, say, implying
that $C$ continues to encircle a degeneracy. 

The rest is plain sailing. Consider $\bham$ in the form (\ref{Htri}),
and let $u_x$ and $u_z$ be one of the points (\ref{uxuz}). As noted
before, $\bham$ divides into two blocks at this point.
Consider now how the
eigenvalues of each block change as $\lam$ is varied. Suppose the
variation is as depicted in \fno{badvar}. This would imply that the total
diabolicity of our system changes discontinuosly, in violation of
the result just proved. Thus a behavior as in \fno{badvar} is impossible,
and the correct picture is as in \fno{okvar}. One can also see that
the behavior can not be that as in \fno{badvar} for some $(m_0,n_0)$ and
that as in \fno{okvar} for others, since there is then no way to keep the
total diabolicity constant.

We thus conclude that the multiplicity $f(m_0,n_0)$ of
the diabolical point at $(m_0,n_0)$ is independent of $\lam$,
and we may find it by evaluating it for any one value of $\lam$.
We do this for $\lam = 0$, as $\bham$ is then not just tridiagonal, but
diagonal. Degeneracy occurs whenever
\beq
w_{m_1} = w_{m_2}, \quad (m_1 \le m_0;\ \ m_2 \ge m_0 + 1).
  \label{eqdiag}
\eeq
Since $w_m = m^2 -(m_0 - n_0)m$ for $\lam = 0$, this condition becomes
\beq
m_1 + m_2 = m_0 - n_0. \label{m1m2}
\eeq
The problem is now one of counting. We leave it as an exercise to show
that with the conditions in \eno{eqdiag} on $m_1$ and $m_2$, and
in Eqs.~(\ref{uxz+}) and (\ref{uxz-}) on $m_0$ and $n_0$, the number
of solutions to \eno{m1m2} is given by
\beq
f(m_0,n_0) = \hf \bigl[ 2J + 1 - |m_0 - n_0|
                               - |m_0 + n_0 + 1|\bigr].
   \label{fmn}
\eeq
This is precisely the multiplicity
implied by \etwo{plhz}{plhx}. One way to see this is to evaluate
$f$ on the rhombi drawn in \fno{dpts}, in particular, the segment lying
in the first quadrant in the $H_x$-$H_z$ plane. Regarding the outermost
rhombus as the first, the next as the second, etc., we have
$m_0 = -(J+1) + k$ on the $k$th one. Further, for the part in the first
quadrant, $n_0$ ranges from $m_0$ to $J - k$. Thus $(m_0 - n_0)$ and
$(m_0 + n_0 +1)$ are both negative, and
\bea
f(m_0,n_0) &=& \hf \bigl[ 2J + 1 + (m_0 - n_0)
                               + (m_0 + n_0 + 1)\bigr] \nnu\\
           &=& J + m_0 + 1 \nnu \\
           &=& k, \label{feqk}
\eea
exactly as expected.

For completenes, we conclude by observing that our results obey
the symmetries of the Hamiltonian. First, for half-integral
$J$, the point at $\bH =0$ corresponds to $m_0 = n_0 = -1/2$. The
value of $f$ is then $J + \tshf$, i.e., every state is doubly
degenerate, as required by Kramers' theorem. Second, the reflection
symmetries $H_x\to -H_x$, $H_z \to - H_z$ are clearly obeyed by the
set (\ref{plhz}) and ({\ref{plhx}). In terms of the quantities $m_0$
and $n_0$, these symmetries correspond to $m_0 \to -(m_0 + 1)$ and
$n_0 \to -(n_0 + 1)$, and it is easy to see from \eno{fmn} that
these leave $f$ unchanged. Lastly, the Hamiltonian (\ref{ham}) has
the following {\it duality} property.  Showing its dependence on $\lam$,
$H_x$ and $H_z$ explicitly by writing $\ham(\lam,H_x,H_z)$, we
note that a $90^{\circ}$ rotation about $(\xhat + \zhat)/\sqrt{2}$
yields the transformation
\beq
\ham(\lam,H_x,H_z) \leftrightarrow -\ham(1-\lam,H_z,H_x).
                         \label{dual}
\eeq
In particular, the spectra of the two Hamiltonians are so related, and
ranking the levels is order of increasing energy, we see that
if the levels with ordinal numbers $i$ and $i+1$ are degenerate when
$H_x = f_x(\lam)$ and $H_y = f_y(\lam)$, where the functions $f_x$ and
$f_y$ are unknown, then level numbers $2J+2 -i$ and $2J+1-i$ are
degenerate when $H_x = f_y(1-\lam)$ and $H_y = f_x(1-\lam)$. This
property is also obeyed by \etwo{plhz}{plhx}, and (\ref{fmn}).

\acknowledgments
AG's research is supported by the NSF via grant number DMR-9616749,
and he is much indebted to Jacques Villain and Wolfgang Wernsdorfer for
useful discussions and correspondence.

\begin{figure}
\caption{Diabolical points for the model Hamiltonian (\ref{ham}), shown
for the case $J=7/2$.  The points can be organized into rhombi (including
the degenerate central rhombus with only one point). All points on a
rhombus have the same multiplicity, i.e., number of simultaneously
degenerate levels. These numbers are as shown. The dashed lines are
self-dual under the transformation (\ref{dual}).}
\label{dpts}
\end{figure}

\begin{figure}
\caption{Impermissible variation of the energy levels with $\lam$ for
fixed $m_0$ and $n_0$. The solid curves are the energies of one
of the blocks of $\bham$, and the dashed curves of the other.}
\label{badvar}
\end{figure}

\begin{figure}
\caption{Same as \fno{badvar}, except that the variation is now
permissible. The solid and dashed curves drawn right next to each
other are in fact coincident.}
\label{okvar}
\end{figure}


\begin{references}
\bibitem{fn1} The remark about ``isolated points" should be
qualified. As noted in \rno{va}, when the dimensionality of the
parameter space is larger than 2 for real, or 3 for complex Hamiltonians,
the precise statement of the theorem is that the {\it codimension} of
the space of a double degeneracy is 2 or 3, as the case may be.
\bibitem{fot} We ignore higher order terms such as $J_{\pm}^4$, that
are necessary to fit the data in \cno{ws} quantitatively.
\end{references}
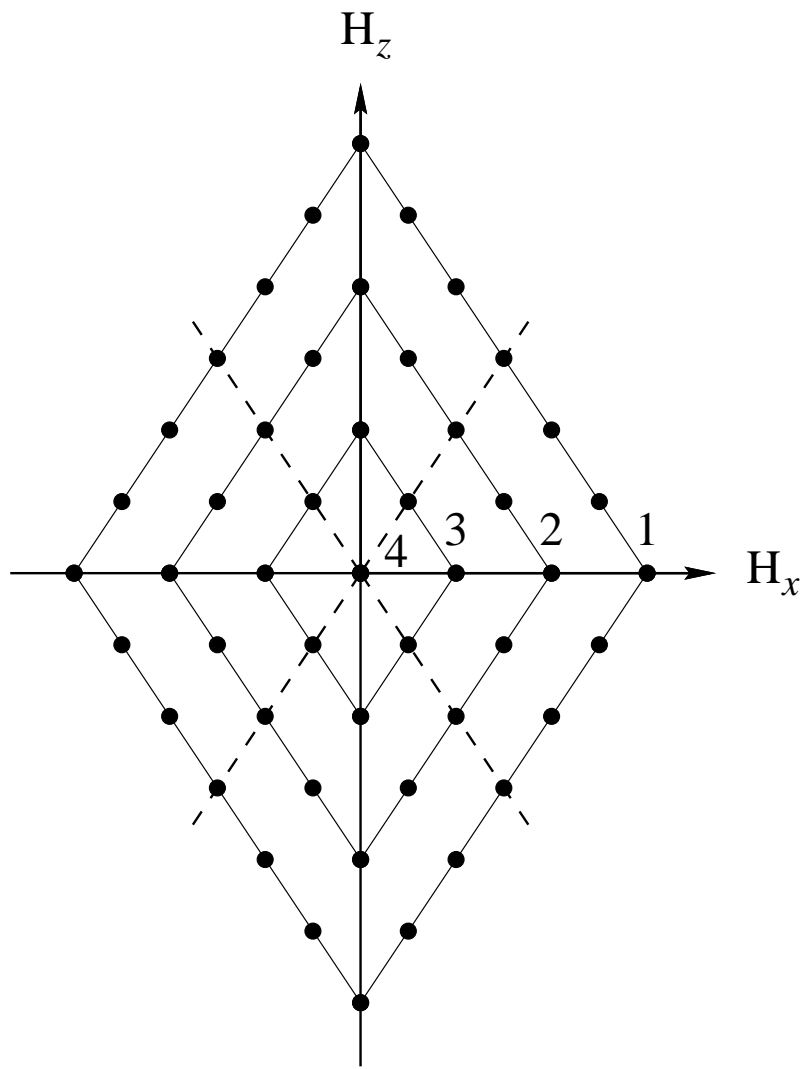
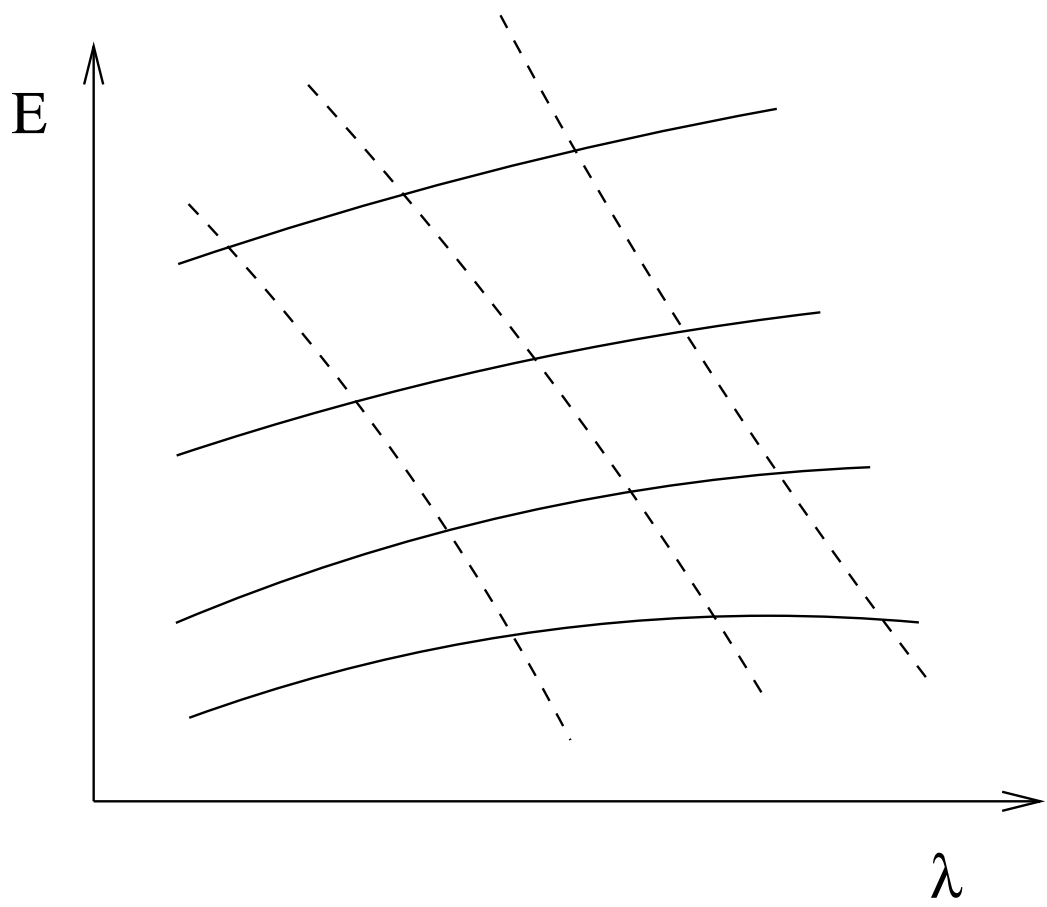
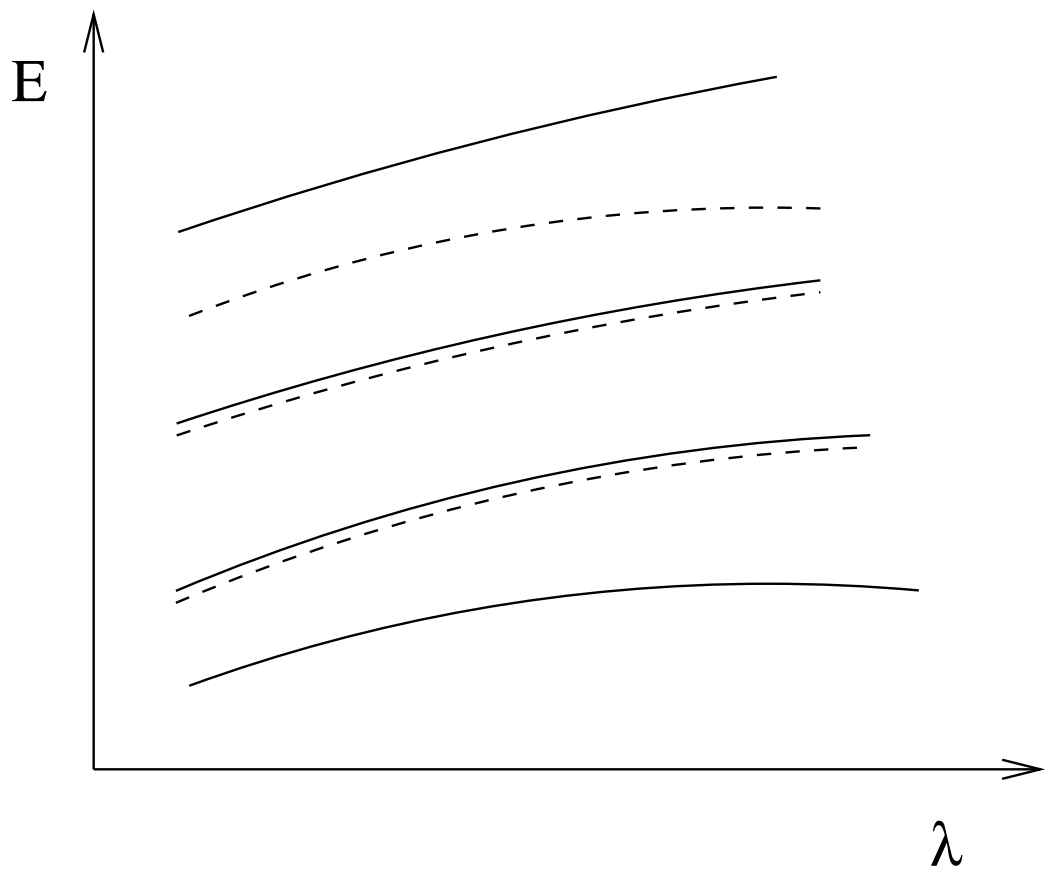
\end{document}